\documentclass[preprint,showpacs,preprintnumbers,amsmath,amssymb,eqsecnum]{revtex4}
\usepackage{graphicx}
\usepackage{dcolumn}
\usepackage{bm}

\newcommand{\bn}{\mbox{\boldmath $n$}}
\newcommand{\ba}{\mbox{\boldmath $a$}}
\newcommand{\bb}{\mbox{\boldmath $b$}}
\newcommand{\bc}{\mbox{\boldmath $c$}}

\newcommand{\be}{\mbox{\boldmath $e$}}
\newcommand{\bff}{\mbox{\boldmath $f$}}
\newcommand{\bp}{\mbox{\boldmath $p$}}
\newcommand{\bq}{\mbox{\boldmath $q$}}
\newcommand{\br}{\mbox{\boldmath $r$}}
\newcommand{\bA}{\mbox{\boldmath $A$}}
\newcommand{\bB}{\mbox{\boldmath $B$}}

\newcommand{\bD}{\mbox{\boldmath $D$}}
\newcommand{\bE}{\mbox{\boldmath $E$}}
\newcommand{\bF}{\mbox{\boldmath $F$}}
\newcommand{\bQ}{\mbox{\boldmath $Q$}}
\newcommand{\bR}{\mbox{\boldmath $R$}}
\newcommand{\bg}{\mbox{\boldmath $g$}}
\newcommand{\bh}{\mbox{\boldmath $h$}}
\newcommand{\bi}{\mbox{\boldmath $i$}}
\newcommand{\bj}{\mbox{\boldmath $j$}}
\newcommand{\bk}{\mbox{\boldmath $k$}}
\newcommand{\bl}{\mbox{\boldmath $l$}}
\newcommand{\bmm}{\mbox{\boldmath $m$}}
\begin{document}



\title{Elliptic solutions of the Skyrme Model\\}

\author{Minoru Hirayama}
 \email{hirayama@sci.toyama-u.ac.jp}
\author{Chang-Guang Shi}
 \email{shicg@jodo.sci.toyama-u.ac.jp}
\author{Jun Yamashita}
\email{fxjun@jodo.sci.toyama-u.ac.jp}
\affiliation{Department of Physics, Toyama University, Gofuku 3190, Toyama 930-8555, {\bf Japan}\\}


\date{March, 17, 2003}

\begin{abstract}
A class of exact solutions of the Skyrme model is obtained. The solutions are described by the Weierstrass $\wp$ function or the Jacobi elliptic function. They are not solitonic but of wave character. They supply us with examples of the superposition of three plane waves in the Skyrme model.

\end{abstract}


\maketitle


\section{\label{sec:Introduction}Introduction\protect\\}

The Skyrme model \cite{Skyrme} has been discussed for more than 40 years. It is an effective field theory describing hadrons \cite{Witten, Jackson}. It is defined by the Lagrangian density
\begin{align}
{\cal L}_S=&-4c_2 \mbox{tr}\left[(g^{\dagger}\partial_{\mu}g)(g^{\dagger}\partial^{\mu}g)\right]+\frac{c_4}{2}\mbox{tr}\left([g^{\dagger}\partial_{\mu}g\,,\,g^{\dagger}\partial_{\nu}g][g^{\dagger}\partial^{\mu}g\,,\,g^{\dagger}\partial^{\nu}g]\right),
\label{eqn:Lagrangian}
\end{align}
where $g(x)$ is an element of $SU$(2) and $c_2$ and $c_4$ are coupling constants. If we define $A_{\mu}^{\alpha}(x)$ and $H_{\mu\nu}^{\alpha}(x)$ by
\begin{align}
&A_{\mu}^{\alpha}=\frac{1}{2i}\mbox{tr}\left(\tau^{\alpha}g^{\dagger}\partial_{\mu}g\right), \label{eqn:Amug} \\
&H_{\mu\nu}^{\alpha}=\partial_{\mu}A_{\nu}^{\alpha}-\partial_{\nu}A_{\mu}^{\alpha},
\end{align}
${\cal L}_S$ is expressed as
\begin{align}
{\cal L}_S=8c_2A_{\mu}^{\alpha}A^{\alpha,\mu}-c_4H_{\mu\nu}^{\alpha}H^{\alpha,\mu\nu},
\label{eqn:LagrangianA}
\end{align}
where $\tau_{\alpha} (\alpha=1, 2, 3)$ are Pauli matrices. The field equation is given as the conservation law
\begin{align}
\partial_{\mu}J^{\alpha, \mu}=0,
\label{eqn:conservationlow}
\end{align}
where $J^{\alpha, \mu}$ $(\alpha=1, 2, 3,\hspace{2mm}\mu=0, 1, 2, 3)$ are defined by
\begin{align}
J^{\alpha, \mu}=2c_2A^{\alpha, \mu}+c_4\varepsilon^{\alpha\beta\gamma}H^{\beta, \mu\nu}A_{\nu}^{\gamma}.
\label{eqn:Jmu}
\end{align}
The field $J^{\alpha, \mu}(x)$ is proportional to the isospin current of the model. Another important conserved current is the baryon number current \cite{Skyrme}:
\begin{align}
N^{\lambda}=\frac{1}{12\pi^2}\varepsilon^{\lambda\mu\nu\rho}\varepsilon^{\alpha\beta\gamma}A_{\mu}^{\alpha}A_{\nu}^{\beta}A_{\rho}^{\gamma}.
\label{eqn:BaryonN}
\end{align}
The conservation law $\partial_{\lambda}N^{\lambda}=0$ follows solely from the definition of $N^{\lambda}(x)$ irrespective of the field equation for $A_{\mu}^{\alpha}(x)$.

By definition, $A_{\mu}^{\alpha}(x)$ satisfies the condition
\begin{align}
\partial_{\mu}A_{\nu}^{\alpha}-\partial_{\nu}A_{\mu}^{\alpha}=2\varepsilon^{\alpha\beta\gamma}A_{\mu}^{\beta}A_{\nu}^{\gamma}.
\label{eqn:conditiona}
\end{align}
It was discussed \cite{Cho} that the Skyrme model is intimately related to the Faddeev model \cite{FaddeevNiemi} defined by
\begin{align}
{\cal L}_F=c_2(\partial_{\mu}\bn)\cdot(\partial^{\mu}\bn)-2c_4F_{\mu\nu}F^{\mu\nu},
\end{align}
where $\bn(x)=\left(n^1(x), n^2(x), n^3(x)\right)$ is a three-component scalar field satisfying
\begin{align}
\bn^2=n^{\alpha}n^{\alpha}=1
\end{align}
and $F_{\mu\nu}(x)$ is given by
\begin{align}
F_{\mu\nu}=\frac{1}{2}\bn\cdot\left(\partial_{\mu}\bn\times\partial_{\nu}\bn\right).
\end{align}
The field equation for $\bn(x)$ is given by
\begin{align}
\partial_{\mu}\left(c_2\bn\times\partial^{\mu}\bn-2c_4F^{\mu\nu}\partial_{\nu}\bn\right)=0.
\label{eqn:equationn}
\end{align}
The Faddeev model is expected to describe the low energy gluonic dynamics of QCD \cite{FaddeevNiemi}. If we define $\bA_{\mu}(x)$ and $\bB_{\mu}(x)$ by $\bA_{\mu}=\left(A_{\mu}^1, A_{\mu}^2, A_{\mu}^3\right), \bB_{\mu}=\left(B_{\mu}^1, B_{\mu}^2, B_{\mu}^3\right)$ with
\begin{align}
B_{\mu}^{\alpha}=\varepsilon^{\alpha\beta\gamma}n^{\beta}\partial_{\mu}n^{\gamma},
\label{eqn:defbb}
\end{align}
Eqs. (\ref{eqn:conservationlow}), (\ref{eqn:conditiona}), and (\ref{eqn:equationn}) become \cite{Hirayama}
\begin{align}
&\partial_{\mu}\left[c_2\bA^{\mu}+c_4\left(\bA^{\mu}\times\bA^{\nu}\right)\times\bA_{\nu}\right]=0, \label{eqn:MotionEQ}\\
&\partial_{\mu}\bA_{\nu}-\partial_{\nu}\bA_{\mu}=2\bA_{\mu}\times\bA_{\nu}, \\
&\partial_{\mu}\left[c_2\bB^{\mu}+c_4\left(\bB^{\mu}\times\bB^{\nu}\right)\times\bB_{\nu}\right]=0.
\end{align}
From the definition (\ref{eqn:defbb}), we have the condition
\begin{align}
\partial_{\mu}\bB_{\nu}-\partial_{\nu}\bB_{\mu}=2\bB_{\mu}\times\bB_{\nu}
\end{align} 
for $\bB_{\mu}$. The parallelism of $\bB_{\mu}$ with $\bA_{\mu}$ is evident. Of course, the two models are different from each other because the degree of freedom of the Skyrme model is 3 while that of the Faddeev model is 2. It can be seen, however, that the configuration $\bB_{\mu}$ yields a restricted class of the configuration $\bA_{\mu}$.

The soliton solutions of the Skyrme model are classified by the baryon number $N=\int d^3V N^0(x)$ and are identified with baryons \cite{Skyrme}. On the other hand, the soliton solutions of the Faddeev model are classified by the topological number $Q_H$ called the Hopf charge and are identified with glueballs \cite{FaddeevNiemi, GH, Hietarinta}. The numerical analysis of these models revealed rich spectra of solitons : knot solitons for the Faddeev model and polyhedral solitons for the Skyrme model \cite{Battye, Sutcliffe}. For example, the solution of the Faddeev model with $Q_H=7$ is the soliton of the trefoil knot structure.

As for the analytic solutions for these models, only a few simple examples are known. Skyrme \cite{Skyrme} found that the configuration 
\begin{align}
g(x)=h(k\cdot x)
\label{eqn:kf}
\end{align}
with $k\cdot x=k_{\mu}x^{\mu}$ leading to the field $A_{\mu}^{\alpha}(x)$ of the form $A_{\mu}^{\alpha}(x)=k_{\mu}f^{\alpha}(k\cdot x)$ satisfies the field equation if $k_{\mu}$ is light-like: $k^2=0$. In a recent paper \cite{Hirayama}, two of the present authors (M.H. and J.Y.) obtained a solution of the form
\begin{align}
g(x)=h(k\cdot x,\hspace{2mm} l\cdot x)
\label{eqn:kl}
\end{align}
with $k^2=l^2=0, \hspace{2mm}k\cdot l\neq 0$. Although the solution (\ref{eqn:kf}) is independent of the coupling constants $c_2$ and $c_4$ in ${\cal L}_S$, the solution (\ref{eqn:kl}) depends on them through the dimensionless combination
\begin{align}
\sigma=\frac{c_2}{c_4(k\cdot l)}.
\end{align}
In this paper, we seek a solution of the form
\begin{align}
g(x)=h(k\cdot x,\hspace{2mm}l\cdot x,\hspace{2mm}m\cdot x)
\label{eqn:klm}
\end{align}
with
\begin{align}
k^2=l^2=m^2=0, \hspace{3mm}k\cdot l,\hspace{2mm}l\cdot m,\hspace{2mm}m\cdot k >0.
\end{align}
It turns out that the solution can be described with the help of the function
\begin{align}
K(\omega)&=\wp(\omega+\omega_3) \nonumber \\
&=e_3+\frac{(e_3-e_1)(e_3-e_2)}{\wp(\omega)-e_3} \label{eqn:Komega}\\
&=e_3+(e_2-e_3)\mbox{sn}^2\left(\sqrt{e_1-e_3}\hspace{1mm}\omega, \,\sqrt{\frac{e_2-e_3}{e_1-e_3}}\right), \nonumber
\end{align}
where $\wp(z)$ is the Weierstrass $\wp$ function satisfying the differential equation
\begin{align}
\left[\wp^{\prime}(z)\right]^2=4\left[\wp(z)-e_1\right]\left[\wp(z)-e_2\right]\left[\wp(z)-e_3\right].
\label{eqn:wpfunc}
\end{align}
Here $e_1, e_2$ and $e_3$ are real constants satisfying $e_1>e_2>e_3$, sn$(u, k)$ is the Jacobi elliptic function of $u$ with the modulus $k$, $2\omega_3$ is the second fundamental period of $\wp(z)$, and $\omega=L\cdot x$ is a linear combination of $k\cdot x, \hspace{2mm}l\cdot x$ and $m\cdot x$. We note that $L^2$ is equal to $c_2/c_4$ multiplied by a constant independent of the momenta $k$, $l$ and $m$. In contrast with the solutions (\ref{eqn:kf}) and (\ref{eqn:kl}) with the vanishing baryon number density, our solution (\ref{eqn:klm}) possesses the nonvanishing baryon number density
\begin{align}
N_0(x)=n_0\left(\frac{c_2}{c_4}\right)^{\frac{3}{2}}\frac{\bk\cdot(\bl\times\bmm)}{\sqrt{(k\cdot l)(l\cdot m)(m\cdot k)}}\frac{d\wp(\omega+\omega_3)}{d\omega},
\label{eqn:BaryonNform}
\end{align}
where $n_0$ is a constant independent of $c_2, c_4, k, l$ and $m$. 

This paper is organized as follows. In Sec.\ref{sec:Formulation of the problem}, we seek a convenient set of variables and fields and write the field equation in a compact way. In Sec.III, we introduce a set of {\it Ans\"atze} for $\bA_{\mu}$ compatible with the field equation. By the above {\it Ans\"atze}, the non linear partial differential equations for $\bA_{\mu}$ are reduced to some algebraic constraints for parameters specifying solutions. Some examples of the solutions of the above algebraic constraints are given. Section V is devoted to a summary and discussion.


\section{\label{sec:Formulation of the problem}Formulation of the problem\protect\\}

In order to discuss the case of $h(k\cdot x, l\cdot x, m\cdot x)$, we introduce the parameters 
\begin{align}
\kappa^i&=\sqrt{\frac{c_4}{c_2}\frac{(k^i\cdot k^j)(k^i\cdot k^k)}{(k^j\cdot k^k)}},
\label{eqn:par}
\end{align}
where the triplet $(i,j,k)$ should be (1,2,3) or (2,3,1) or (3,1,2) and $k^1=k$, $k^2=l$, $k^3=m$.
Then we have
\begin{align}
\frac{k^i\cdot k^j}{\kappa^i \kappa^j}=\left\{\begin{array}{ll}
0&, \quad i=j,\\
\displaystyle{\frac{c_2}{c_4}}&, \quad i\neq j.
\end{array}\right.
\label{eqn:kap}
\end{align}
We now introduce
\begin{equation}
\xi^i=\frac{k^i\cdot x}{\kappa^i}\hspace{5mm}(i=1,2,3)
\label{eqn:xi}
\end{equation}
and write $\xi=\xi^1$, $\eta=\xi^2$ and $\zeta=\xi^3$. For $g(x)=g(\xi^1,\xi^2,\xi^3)$, $A_{\mu}^\alpha(x)$ in Eq. (\ref{eqn:Amug}) can be written as
\begin{align}
A_{\mu}^\alpha&=\sum_{i=1}^3\frac{{k_\mu}^i}{\kappa^i}a_i^\alpha,\nonumber\\
a_i^\alpha&=\frac{1}{2i} \text{tr}\left(\tau^\alpha g^\dagger\frac{\partial g}{\partial\xi^i}\right).
\label{eqn:Amu}
\end{align} 
If we define $\ba$, $\bb$ and $\bc$ by 
\begin{align}
\ba=(a_1^1,a_1^2,a_1^3), \hspace{5mm}\bb=(a_2^1,a_2^2,a_2^3), \hspace{5mm}\bc=(a_3^1,a_3^2,a_3^3),
\label{eqn:aia}
\end{align} 
$\bA_{\mu}(x)$ can be rewritten as
\begin{equation}
\bA_{\mu}=\frac{k_{\mu}\ba}{\kappa^1}+\frac{l_{\mu}\bb}{\kappa^2}+\frac{m_{\mu}\bc}{\kappa^3}.
\label{eqn:amur}
\end{equation}
We find that the integrability condition (\ref{eqn:conditiona}) becomes
\begin{align}
&\frac{l_{\mu}k_{\nu}-l_{\nu}k_{\mu}}{\kappa^1 \kappa^2}\left(\frac{\partial\ba}{\partial\eta}-\frac{\partial\bb}{\partial\xi}-2\bb\times\ba\right)+\frac{m_{\mu}l_{\nu}-m_{\nu}l_{\mu}}{\kappa^2 \kappa^3}\left(\frac{\partial\bb}{\partial\zeta}-\frac{\partial\bc}{\partial\eta}-2\bc\times\bb\right) \nonumber\\
&+\frac{m_{\mu}k_{\nu}-m_{\nu}k_{\mu}}{\kappa^3 \kappa^1}\left(\frac{\partial\ba}{\partial\zeta}-\frac{\partial\bc}{\partial\xi}-2\bc\times\ba\right)=0,
\label{eqn:equ}
\end{align}
which yields
\begin{align}\begin{array}{ll}
\displaystyle{\frac{\partial\ba}{\partial\eta}-\frac{\partial\bb}{\partial\xi}}&=2(\bb\times\ba),\\
\displaystyle{\frac{\partial\ba}{\partial\zeta}-\frac{\partial\bc}{\partial\xi}}&=2(\bc\times\ba),\\
\displaystyle{\frac{\partial\bb}{\partial\zeta}-\frac{\partial\bc}{\partial\eta}}&=2(\bc\times\bb).
\end{array}\label{eqn:condition}
\end{align}
The second term on the right hand side of Eq. (\ref{eqn:Jmu}) is
expressed as
\begin{align}
&\varepsilon^{\alpha\beta\gamma}H^{\beta}_{\mu\nu}A^{\gamma,\nu}=\frac{2}{\kappa^1\kappa^2\kappa^3}\left[(l\cdot m)k_{\mu}D^{\alpha}+(m\cdot k)l_{\mu}E^{\alpha}+(l\cdot k)m_{\mu}F^{\alpha}\right], 
\label{eqn:lhs}
\end{align}
where $\bD$, $\bE$, $\bF$ and $\alpha$ are given by 
\begin{align}
\bD&=\ba\times(\bb\times\ba)+\ba\times(\bc\times\ba)+\bc\times(\bb\times\ba)+\bb\times(\bc\times\ba),\nonumber\\
\bE&=\bb\times(\ba\times\bb)+\bb\times(\bc\times\bb)+\bc\times(\ba\times\bb)+\ba\times(\bc\times\bb), \label{eqn:DEF}\\
\bF&=\bc\times(\ba\times\bc)+\bc\times(\bb\times\bc)+\ba\times(\bb\times\bc)+\bb\times(\ba\times\bc).\nonumber
\end{align}
Therefore, the field equation (\ref{eqn:conservationlow}) becomes
\begin{align}
&\left(\frac{\partial}{\partial\eta}+\frac{\partial}{\partial\zeta}\right)\left(\ba+\bD\right)+\left(\frac{\partial}{\partial\xi}+\frac{\partial}{\partial\zeta}\right)\left(\bb+\bE\right)+\left(\frac{\partial}{\partial\xi}+\frac{\partial}{\partial\eta}\right)\left(\bc+\bF\right)=0.
\label{eqn:EQMotion1}
\end{align}    
It is a vector nonlinear differential equation. It is difficult to discuss it generally, so we will introduce some {\it Ans\"atze} in the next section.


\section{\label{sec:Ansatzs}Ans\"atze\protect\\}

\subsection{\label{sec:Parameters specifying the Solution}Parameters specifying the solution\protect\\}

In order to obtain the solutions of Eqs. (\ref{eqn:condition}) and (\ref{eqn:EQMotion1}), we first assume
\begin{align}
\frac{\partial\ba_i}{\partial\xi^j}=C_{ij}\ba_i\times\ba_j \hspace{3mm}(i\neq j),
\label{eqn:ansatz1}
\end{align}
with $\ba_1=\ba,\hspace{2mm}\ba_2=\bb,$ and $\hspace{2mm}\ba_3=\bc$. Here $C_{ij}$ are constants. Since the condition (\ref{eqn:condition}) can be written as
\begin{align}
C_{ij}+C_{ji}=-2 \hspace{5mm} (ij=12, 23, 31),
\end{align}
we are led to the parametrization
\begin{align}
&C_{23}=\alpha, \hspace{2mm} C_{31}=\beta, \hspace{2mm} C_{12}=\gamma, \nonumber \\
&C_{32}=-2-\alpha, \hspace{2mm}C_{13}=-2-\beta,\hspace{2mm} C_{21}=-2-\gamma.
\end{align} 
It is interesting to note that the three conditions
\begin{align}
\frac{\partial}{\partial\xi^k}\left(\frac{\partial\ba_i}{\partial\xi^j}\right)=\frac{\partial}{\partial\xi^j}\left(\frac{\partial\ba_i}{\partial\xi^k}\right)
\label{eqn:dcondition}
\end{align}
for $i\neq j,\hspace{2mm} j\neq k$ and $k\neq i$ are satisfied if $\alpha, \beta$ and $\gamma$ satisfy the single relation
\begin{align}
\alpha\beta\gamma=(\alpha+2)(\beta+2)(\gamma+2).
\label{eqn:constraints1}
\end{align}
In addition to Eq. (\ref{eqn:ansatz1}), we assume
\begin{align}
&\frac{\partial\ba}{\partial\xi}=\kappa\left(\ba\times\bc\right)+\lambda\left(\ba\times\bb\right), \nonumber\\
&\frac{\partial\bb}{\partial\eta}=\rho\left(\bb\times\ba\right)+\sigma\left(\bb\times\bc\right), \label{eqn:ansatz2} \\
&\frac{\partial\bc}{\partial\zeta}=\nu\left(\bc\times\bb\right)+\mu\left(\bc\times\ba\right), \nonumber
\end{align}
where $\kappa, \lambda, \rho, \sigma, \nu$ and $\mu$ are constants. Then, the conditions (\ref{eqn:dcondition}) yield
\begin{align}
&\alpha(\alpha+2)+\nu\sigma=0, &\beta(\beta+2)+\mu\kappa=0, \nonumber\\
&\gamma(\gamma+2)+\lambda\rho=0, &\lambda(\alpha+\beta+2)-\kappa\nu=0, \label{eqn:constraints7} \\
&\sigma(\beta+\gamma+2)-\rho\kappa=0, &\mu(\alpha+\gamma+2)-\nu\rho=0, \nonumber\\
&\kappa\nu\rho+\lambda\mu\sigma=0. & \nonumber
\end{align}
It can be seen that the eight conditions (\ref{eqn:constraints1}) and (\ref{eqn:constraints7}) for nine parameters are equivalent to five conditions
\begin{align}
&\alpha\beta\gamma=(\alpha+2)(\beta+2)(\gamma+2), \nonumber\\
&\alpha(\alpha+2)+\nu\sigma=0, \nonumber\\
&\beta(\beta+2)+\mu\kappa=0, \label{eqn:constraints5}\\
&\gamma(\gamma+2)+\lambda\rho=0, \nonumber\\
&\kappa\nu\rho+\lambda\mu\sigma=0. \nonumber
\end{align}
We now regard the independent parameters as $\alpha, \beta, \mu$ and $\nu$. The other parameters $\gamma, \kappa, \lambda, \rho$ and $\sigma$ are given as
\begin{align}
&\kappa=-\frac{\beta(\beta+2)}{\mu}, \nonumber\\
&\sigma=-\frac{\alpha(\alpha+2)}{\nu}, \nonumber\\
&\lambda=-\frac{\nu}{\mu}\frac{\beta(\beta+2)}{\alpha+\beta+2}, \\
&\rho=\frac{\mu}{\nu}\frac{\alpha(\alpha+2)}{\alpha+\beta+2}, \nonumber\\
&\gamma=-\frac{(\alpha+2)(\beta+2)}{\alpha+\beta+2}. \nonumber
\label{eqn:relation5}
\end{align}

\subsection{\label{sec:Consequenes of the Ansatz}Consequences of the Ans\"atze\protect\\}

Our {\it Ans\"atze} proposed in the previous subsection readily lead us to the consequence that $\ba^2, \bb^2$ and  $\bc^2$ are constant. We also find the relations
\begin{align}
-\frac{1}{\kappa}\frac{\partial(\ba\cdot\bb)}{\partial\xi}&=\frac{1}{\sigma}\frac{\partial(\ba\cdot\bb)}{\partial\eta}=\frac{\lambda}{\kappa\nu}\frac{\partial(\ba\cdot\bb)}{\partial\zeta} \nonumber\\
&=\frac{\sigma}{\rho\kappa}\frac{\partial(\bb\cdot\bc)}{\partial\xi}=-\frac{1}{\rho}\frac{\partial(\bb\cdot\bc)}{\partial\eta}=\frac{1}{\mu}\frac{\partial(\bb\cdot\bc)}{\partial\zeta} \\
&=\frac{1}{\lambda}\frac{\partial(\bc\cdot\ba)}{\partial\xi}=\frac{\mu}{\nu\rho}\frac{\partial(\bc\cdot\ba)}{\partial\eta}=-\frac{1}{\nu}\frac{\partial(\bc\cdot\ba)}{\partial\zeta}\nonumber\\
&=\ba\cdot\left(\bb\times\bc\right).\nonumber
\label{eqn:adotbc}
\end{align}
From the consistency conditions
\begin{align}
\frac{\partial}{\partial\xi^i}\left(\frac{\partial A}{\partial\xi^j}\right)=\frac{\partial}{\partial\xi^j}\left(\frac{\partial A}{\partial\xi^i}\right) \hspace{3mm} (i,j=1, 2, 3, A=\ba\cdot\bb,\hspace{3mm}\bb\cdot\bc,\hspace{3mm}\bc\cdot\ba),
\end{align}
we obtain
\begin{align}
\kappa\frac{\partial\left[\ba\cdot(\bb\times\bc)\right]}{\partial\eta}+\sigma\frac{\partial\left[\ba\cdot(\bb\times\bc)\right]}{\partial\xi}&=\rho\frac{\partial\left[\ba\cdot(\bb\times\bc)\right]}{\partial\zeta}+\mu\frac{\partial\left[\ba\cdot(\bb\times\bc)\right]}{\partial\eta} \nonumber\\
&=\nu\frac{\partial\left[\ba\cdot(\bb\times\bc)\right]}{\partial\xi}+\lambda\frac{\partial\left[\ba\cdot(\bb\times\bc)\right]}{\partial\zeta} \nonumber \\
&=0.
\end{align}
With the aid of the relations (\ref{eqn:relation5}) and (\ref{eqn:adotbc}), we conclude that $\ba\cdot\bb,\hspace{2mm}\bb\cdot\bc,\hspace{2mm}\bc\cdot\ba$ and $\ba\cdot(\bb\times\bc)$ are given in the following way:
\begin{align}
&\ba\cdot\bb=(\alpha+\beta+2)J(\omega)+d_1, \nonumber\\
&\bb\cdot\bc=\mu J(\omega)+d_2 \label{eqn:ab}, \\
&\bc\cdot\ba=-\nu J(\omega)+d_3 \nonumber, \\
&\ba\cdot\left(\bb\times\bc\right)=-\alpha(\alpha+2)\frac{\mu}{\nu}\frac{dJ(\omega)}{d\omega}, \label{eqn:p} \\
&\omega=\rho\kappa\xi-\rho\sigma\eta+\mu\sigma\zeta,  \label{eqn:omega}
\end{align}
where $J(\omega)$ is a function of $\omega$ to be determined and $d_1$, $d_2$ and $d_3$ are arbitrary constants. The function $J(\omega)$ is fixed as follows. From the formula
\begin{align}
\left[\ba\cdot(\bb\times\bc)\right]^2=\left|\begin{array}{ccc}\ba^2 &\ba\cdot\bb &\ba\cdot\bc \\
\ba\cdot\bb &\bb^2 &\bb\cdot\bc \\
\ba\cdot\bc &\bb\cdot\bc &\bc^2 \end{array}\right|
\end{align}
and Eqs. (\ref{eqn:ab}) and (\ref{eqn:p}), we observe that $K(\omega)$ defined by
\begin{align}
&K(\omega)=\frac{1}{z_1}J(\omega)-\frac{z_2}{z_1}, \nonumber\\
&z_1=-\frac{2\alpha^2(\alpha+2)^2\mu}{(\alpha+\beta+2)\nu^3}, \\
&z_2=\frac{-\mu^2\ba^2-\nu^2\bb^2-(\alpha+\beta+2)^2\bc^2-2\mu\nu d_1+2(\alpha+\beta+2)(\mu d_3-\nu d_2)}{6(\alpha+\beta+2)\mu\nu} \nonumber
\label{eqn:Jz1z2}
\end{align}
should satisfy an equation of the form
\begin{align}
\left[\frac{dK(\omega)}{d\omega}\right]^2&=4\left[K(\omega)\right]^3-g_2K(\omega)-g_3 \\ 
&=4\left[K(\omega)-e_1\right]\left[K(\omega)-e_2\right]\left[K(\omega)-e_3\right],\nonumber
\end{align}
where the constants $e_1, e_2, e_3, g_2$ and $g_3$ are complicated functions of $\alpha, \beta, \mu, \nu, \ba^2, \bb^2, \bc^2, d_1, d_2$ and $d_3$ and satisfy the relations
\begin{align}
&e_1+e_2+e_3=0, \nonumber\\
&e_2e_3+e_3e_1+e_1e_2=-\frac{1}{4}g_2, \\
&e_1e_2e_3=\frac{1}{4}g_3. \nonumber
\end{align}
This equation indicates that $K(\omega)$ is given by the Weierstrass $\wp$ function as $K(\omega)=\wp(\omega+\mbox{const})$. We hereafter assume that the constants $\alpha, \beta, \mu, \nu, \ba^2, \bb^2, \bc^2, d_1, d_2$ and $d_3$ are chosen so that the inequality
\begin{align}
\left(g_2\right)^3-27\left(g_3\right)^2>0
\label{eqn:wpinequality}
\end{align}
holds. Then the constants $e_1, e_2$ and $e_3$ are real and can be assumed to satisfy $e_1>e_2>e_3$.
To remove the poles for the real values of $\omega$, we adopt $K(\omega)$ given by Eq. (\ref{eqn:Komega}). The explicit formula for $\omega_3$ is given by
\begin{align}
\omega_3=\frac{i}{2}\int_{-\infty}^{e_3}\frac{du}{\sqrt{(e_1-u)(e_2-u)(e_3-u)}}.
\end{align}
We find that $\omega_3$ is a purely imaginary number. Other choices of the integral constant lead to a complex-valued $K(\omega)$. From Eqs. (\ref{eqn:xi}) and (\ref{eqn:omega}), $\omega$ is calculated as
\begin{align}
&\omega=L\cdot x \nonumber \\
&L=\frac{\alpha(\alpha+2)}{\nu^2(\alpha+\beta+2)}\left[-\nu\beta(\beta+2)\frac{k}{\kappa^1}+\mu\alpha(\alpha+2)\frac{l}{\kappa^2}-\mu\nu(\alpha+\beta+2)\frac{m}{\kappa^3}\right]. \label{eqn:L}
\end{align}
As was noted below Eq. (\ref{eqn:wpfunc}), $L^2$ is a constant multiple of $c_2/c_4$:
\begin{align}
L^2=\frac{2\alpha^2(\alpha+2)^2}{(\alpha+\beta+2)^2}\left[-\alpha\beta(\alpha+2)(\beta+2)+\beta(\beta+2)(\alpha+\beta+2)\nu-\alpha(\alpha+2)(\alpha+\beta+2)\mu\right]\frac{\mu}{\nu^3}\frac{c_2}{c_4}.
\end{align}
If we define the differential operator $D$ by
\begin{align}
D=\mu(\alpha+2)\frac{\partial}{\partial\xi}-\nu\beta\frac{\partial}{\partial\eta}-\beta(\alpha+2)\frac{\partial}{\partial\zeta},
\end{align}
$\ba, \bb, \bc$ and $\omega$ satisfy
\begin{align}
D\ba=D\bb=D\bc=D\omega=0.
\label{eqn:DaDbDc}
\end{align}
Since there exists another independent variable $\omega^{\prime}$ satisfying $D\omega^{\prime}=0$, e.g.
\begin{align}
\omega^{\prime}=\frac{\xi}{\mu(\alpha+2)}+\frac{\eta}{\nu\beta},
\label{eqn:omega'}
\end{align}
$\ba, \bb$ and $\bc$ are represented as
\begin{align}
\ba=\ba(\omega, \omega^{\prime}),\hspace{2mm} \bb=\bb(\omega, \omega^{\prime}),\hspace{2mm} \bc=\bc(\omega, \omega^{\prime}). \label{eqn:babbbc}
\end{align}
We have thus seen that the general solution of our {\it Ansatz} are given by Eqs. (\ref{eqn:L}), (\ref{eqn:omega'}) and (\ref{eqn:babbbc}). We note that we have not yet made use of the field equation (\ref{eqn:EQMotion1}). In the next section, we shall see that the field equation can be used to constrain the allowed values of the independent parameters $\alpha, \beta, \mu$ and $\nu$.

\subsection{\label{sec:Baryon number density}Baryon number density\protect\\}

The baryon number current $N^{\lambda}(x)$ in the Skyrme model was defined by Eq.(\ref{eqn:BaryonN}). Substituting Eq. (\ref{eqn:Amu}) into Eq. (\ref{eqn:BaryonN}) , we obtain the baryon number density (\ref{eqn:BaryonNform}) with $n_0$ given by
\begin{align}
n_0=\frac{1}{\pi^2}\frac{\alpha^3(\alpha+2)^3\mu^2}{(\alpha+\beta+2)\nu^4}.
\end{align}


\section{\label{sec:solving the field equation}solving the field equation\protect\\}

To discuss the field equation, it is convenient to introduce $\bR$ and $\bQ$ by
\begin{align}
\bR&=\left(\frac{\partial}{\partial\eta}+\frac{\partial}{\partial\zeta}\right)\ba+\left(\frac{\partial}{\partial\xi}+\frac{\partial}{\partial\zeta}\right)\bb+\left(\frac{\partial}{\partial\xi}+\frac{\partial}{\partial\eta}\right)\bc, \nonumber\\
\bQ&=\frac{\partial}{\partial\zeta}\left(\bD+\bE\right)+\frac{\partial}{\partial\xi}\left(\bE+\bF\right)+\frac{\partial}{\partial\eta}\left(\bD+\bF\right).
\end{align}
Then the field equation (\ref{eqn:EQMotion1}) becomes
\begin{align}
\bR+\bQ=0.
\label{eqn:RQ}
\end{align} 
With the help of the {\it Ans\"atze} (\ref{eqn:ansatz1}) and (\ref{eqn:ansatz2}), $\bR$ and $\bQ$ can be expressed solely by $\ba$, $\bb$ and $\bc$ without their derivatives. After some manipulations, we obtain
\begin{align}
\bR&=2(\gamma+1)(\ba\times\bb)+2(\alpha+1)(\bb\times\bc)+2(\beta+1)(\bc\times\ba),\nonumber\\
\bQ&=\left[\ba\cdot(\bb\times\bc)\right]\left(A\ba+B\bb+C\bc\right)+D(\ba\times\bb)+E(\bb\times\bc)+F(\bc\times\ba),
\end{align}
where $A$, $B$, $C$, $D$, $E$ and $F$ are given by
\begin{align}
A&=\beta+\gamma-2\alpha+\nu+\rho-\mu-\sigma, \nonumber\\
B&=\alpha+\gamma-2\beta+\kappa+\nu-\lambda-\mu, \nonumber\\
C&=\alpha+\beta-2\gamma+\rho+\kappa-\sigma-\lambda, \nonumber\\
D&=-\rho\bm{a}^2+\lambda\bm{b^2}+(2\gamma+2+\lambda-\rho)\bm{c}^2-(2\gamma+2)\bm{a}\cdot\bm{b} \nonumber\\
&+(-2\gamma-2+2\lambda)\bb\cdot\bc-(2\gamma+2+2\rho)\bc\cdot\ba,\\
E&=(2\alpha+2+\sigma-\nu)\ba^2-\nu\bb^2+\sigma\bc^2-(2\alpha+2+2\nu)\ba\cdot\bb \nonumber\\
&-(2\alpha+2)\bb\cdot\bc+(-2\alpha-2+2\sigma)\bc\cdot\ba \nonumber\\
F&=\mu\ba^2+(2\beta+2+\mu-\kappa)\bb^2-\kappa\bc^2+(-2\beta-2+2\mu)\ba\cdot\bb-(2\beta+2\kappa+2)\bb\cdot\bc \nonumber\\
&-(2\beta+2)\bc\cdot\ba. \nonumber
\end{align}
Assuming that $\ba\cdot(\bb\times\bc)$ is nonvanishing, Eq. (\ref{eqn:RQ}) is equivalent to the condition 
\begin{align}
\mathcal{S}=\mathcal{T}=\mathcal{U}=0,
\end{align}
 where $\mathcal{S}$, $\mathcal{T}$, $\mathcal{U}$ are defined by
\begin{align}
\mathcal{S}&=\frac{1}{\ba\cdot(\bb\times\bc)}\left(\ba\cdot(\bR+\bQ)\right), \nonumber\\
\mathcal{T}&=\frac{1}{\ba\cdot(\bb\times\bc)}\left(\bb\cdot(\bR+\bQ)\right), \label{eqn:STU}\\
\mathcal{U}&=\frac{1}{\ba\cdot(\bb\times\bc)}\left(\bc\cdot(\bR+\bQ)\right). \nonumber
\end{align}
By making use of Eq. (\ref{eqn:ab}), we see that $\mathcal{S}$, $\mathcal{T}$ and $\mathcal{U}$ take the following form:
\begin{align}
\mathcal{S}&=sJ(\omega)+s^{\prime}, \nonumber\\
\mathcal{T}&=tJ(\omega)+t^{\prime}, \label{eqn:mastu}\\
\mathcal{U}&=uJ(\omega)+u^{\prime}. \nonumber
\end{align}
where $s$, $t$, $u$, $s^{\prime}$, $t^{\prime}$, $u^{\prime}$ are constant. We now find that the field equation has been reduced to the equations 
\begin{align}
s=t=u=s^{\prime}=t^{\prime}=u^{\prime}=0.
\end{align}

\subsection{\label{subsec:Solving $s'=t'=u'=0$}Solving $s'=t'=u'=0$\protect\\}

We first discuss the equations $s^{\prime}=t^{\prime}=u^{\prime}=0$. The constants $s^{\prime}$, $t^{\prime}$ and $u^{\prime}$ in Eq. (\ref{eqn:mastu}) are expressed as
\begin{align}
s^{\prime}&=Bd_1+Cd_3+s_0, \nonumber\\
t^{\prime}&=Ad_1+Cd_2+t_0, \label{eqn:s't'u'}\\
u^{\prime}&=Bd_2+Ad_3+u_0, \nonumber
\end{align} 
where  $s_0$, $t_0$ and $u_0$ are given by 
\begin{align}
s_0&=(\beta+\gamma+2+\rho-\mu)\ba^2-\nu\bb^2+\sigma\bc^2+2(\alpha+1), \nonumber\\
t_0&=\mu\ba^2+(\alpha+\gamma+2+\nu-\lambda)\bb^2-\kappa\bc^2+2(\beta+1), \label{eqn:s0t0u0}\\
u_0&=-\rho\ba^2+\lambda\bb^2+(\alpha+\beta+2+\kappa-\sigma)\bc^2+2(\gamma+1). \nonumber
\end{align}
To make $s^{\prime}$, $t^{\prime}$ and $u^{\prime}$ vanishing, we have only to choose the arbitrary constants $d_1$, $d_2$ and $d_3$ as 
\begin{align} 
d_1&=\frac{Cu_0-As_0-Bt_0}{2AB}, \nonumber\\
d_2&=\frac{As_0-Bt_0-Cu_0}{2BC}, \label{eqn:c1c2c3}\\
d_3&=\frac{Bt_0-As_0-Cu_0}{2AC}. \nonumber
\end{align}
\subsection{\label{subsec:Solving $s=t=u=0$}{Solving $s=t=u=0$}\protect\\}

We next consider the three equations $s=t=u=0$. It can be seen that they are equivalent to the following algebraic equations for $\mu$ and $\nu$: 
\begin{align}
F_i(\mu,\nu)=p_i\mu^2+2h_i\mu\nu+q_i\nu^2+2g_i\mu+2f_i\nu+r_i=0\quad(i=1,2,3),
\label{eqn:fi}
\end{align} 
where $p_i$ $h_i$, $q_i$, $g_i$, $f_i$ and $r_i$ are given by $\alpha$ and $\beta$ as follows:
\begin{align}&\begin{array}{ll}
p_1&\hspace{-2mm}=(2+2\alpha+\beta)(4+2\alpha+\beta), \\
q_1&\hspace{-2mm}=\beta(2+\beta), \\
r_1&\hspace{-2mm}=\beta(2+\beta)(2+\alpha+\beta)^2, \\
h_1&\hspace{-2mm}=2\alpha(1+\beta)+(2+\beta)^2, \\
g_1&\hspace{-2mm}=(2+\alpha+\beta)\left[2\alpha(1+\beta)+(2+\beta)^2\right], \\
f_1&\hspace{-2mm}=-\beta(2+\beta)(2+\alpha+\beta),
\end{array} \\
&\begin{array}{ll}
p_2&\hspace{-2mm}=\alpha(2+\alpha), \\
q_2&\hspace{-2mm}=(2+\alpha+2\beta)(4+\alpha+2\beta), \\ 
r_2&\hspace{-2mm}=\alpha(2+\alpha)(2+\alpha+\beta)^2, \\
h_2&\hspace{-2mm}=(2+\alpha)^2+2(1+\alpha)\beta, \\
g_2&\hspace{-2mm}=\alpha(2+\alpha)(2+\alpha+\beta), \\
f_2&\hspace{-2mm}=-(2+\alpha+\beta)\left[(2+\alpha)^2+2(1+\alpha)\beta\right],
\end{array} \\
&\begin{array}{ll}
p_3&\hspace{-2mm}=1, \\
q_3&\hspace{-2mm}=1, \\ 
r_3&\hspace{-2mm}=-4+(\alpha-\beta)^2, \\
h_3&\hspace{-2mm}=-1, \\
g_3&\hspace{-2mm}=\beta-\alpha, \\
f_3&\hspace{-2mm}=\beta-\alpha.
\end{array}
\end{align}            
The parameters $\alpha$ and $\beta$ must be chosen so that the simultaneous equations (\ref{eqn:fi}) of second order for $\mu$ and $\nu$ possess a common root $(\mu,\nu)$. To obtain the condition that $\alpha$ and $\beta$ must satisfy, assuming $p_1$, $q_1$, $h_1\neq 0$, we rewrite the equation  $F_i({\mu,\nu})=0\hspace{3mm}(i=1,2,3)$ as follows:
\begin{align}
Q_1\nu^2+2G_1\mu+2F_1\nu+R_1&=0,\label{eqn:Q}\\
Q_2\mu^2+2G_2\mu+2F_2\nu+R_2&=0,\label{eqn:Q'}\\
2Q_3\mu\nu+2G_3\mu+2F_3\nu+R_3&=0,\label{eqn:Q''}
\end{align}
where $Q_i$, $G_i$, $F_i$ and $R_i$ $(i=1, 2, 3)$ are defined by
\begin{align}&\begin{array}{ll}
&Q_1=(\bp\times\bh)\cdot\bq=\varepsilon_{ijk}p_ih_jq_k,\\
&Q_2=(\bq\times\bh)\cdot\bp=-Q_1,\\
&Q_3=(\bp\times\bq)\cdot\bh=-Q_1,
\end{array} \\
&\begin{array}{ll}
&G_1=(\bp\times\bh)\cdot\bg,\\
&G_2=(\bq\times\bh)\cdot\bg,\\
&G_3=(\bp\times\bq)\cdot\bg,
\end{array} \\
&\begin{array}{ll}
&F_1=(\bp\times\bh)\cdot\bff,\\
&F_2=(\bq\times\bh)\cdot\bff,\\
&F_3=(\bp\times\bq)\cdot\bff,
\end{array} \\
&\begin{array}{ll}
&R_1=(\bp\times\bh)\cdot\br,\\
&R_2=(\bq\times\bh)\cdot\br,\\
&R_3=(\bp\times\bq)\cdot\br.
\end{array}
\end{align}
If we eliminate $\mu$ from Eqs. (\ref{eqn:Q}) and (\ref{eqn:Q'}), we obtain
\begin{align}
A_0\nu^4+A_1\nu^3+A_2\nu^2+A_3\nu+A_4=0 \label{eqn:A}
\end{align}
with
\begin{align}\begin{array}{ll}
&A_0=Q_1^2 Q_2, \\
&A_1=4F_1 Q_1 Q_2, \\
&A_2=4F_1^2Q_2+2Q_1Q_2R_1-4Q_1G_1G_2, \\
&A_3=4Q_2F_1R_1-8F_1G_1G_2+8F_2G_1^2, \\
&A_4=Q_2R_1^2-4G_1G_2R_1+4R_2G_1^2.
\end{array}
\end{align}
If we eliminate $\mu$ from Eqs. (\ref{eqn:Q}) and (\ref{eqn:Q''}), we obtain 
\begin{align}
B_0\nu^3+B_1\nu^2+B_2\nu+B_3=0 \label{eqn:B}
\end{align}
with
\begin{align}\begin{array}{ll}
&B_0=Q_1Q_3, \\
&B_1=Q_1G_3+2F_1Q_3, \\
&B_2=R_1Q_3+2F_1G_3-2F_3G_1, \\
&B_3=G_3R_1-G_1R_3.
\end{array}
\end{align}
The condition for the two equations (\ref{eqn:A}) and (\ref{eqn:B}) to have a common root $\nu$ is that the resultant for these equations vanishes:
\begin{align}
R(\alpha,\beta)\equiv\begin{vmatrix}
A_0&A_1&A_2&A_3&A_4&0&0\\
0&A_0&A_1&A_2&A_3&A_4&0\\
0&0&A_0&A_1&A_2&A_3&A_4\\
B_0&B_1&B_2&B_3&0&0&0\\
0&B_0&B_1&B_2&B_3&0&0\\
0&0&B_0&B_1&B_2&B_3&0\\
0&0&0&B_0&B_1&B_2&B_3\\
\end{vmatrix}=0.
\label{eqn:resultant}
\end{align}
We note that the condition (\ref{eqn:resultant}) is necessary but not sufficient for the original equations (\ref{eqn:fi}) to have a common root $(\mu,\nu)$.  In general, the condition (\ref{eqn:resultant}) is a very complicated relation between $\alpha$ and $\beta$. It is clear, however, that there exist infinite number of pairs $(\alpha,\beta)$ that satisfy (\ref{eqn:resultant}). 

For simplicity, let us consider the case $\alpha=\beta$. Then Eq. (\ref{eqn:resultant}) becomes
\begin{align}
\beta^4(\beta+1)^{56}(\beta+2)^4R_1(\beta)R_2(\beta)=0,
\end{align}
where $R_1(\beta)$ and $R_2(\beta)$ are defined by
\begin{align}
&R_1(\beta)=\beta^4+8\beta^3+18\beta^2+16\beta+4,\nonumber\\
&R_2(\beta)=\beta^4-6\beta^2-8\beta-4.
\end{align} 
We here exclude the case $\alpha\beta(\alpha+2)(\beta+2)(\alpha+\beta+2)=0$, in which our formulation above should be modified. We find two real roots 
\begin{align}
\beta_1&=-2-\frac{1}{\sqrt{2}}-\frac{\sqrt{5+4\sqrt{2}}}{2},\nonumber\\ 
\beta_2&=-2-\frac{1}{\sqrt{2}}+\frac{\sqrt{5+4\sqrt{2}}}{2} 
\end{align}
of $R_1(\beta)=0$ and two real roots 
\begin{align}
\beta_3&=\frac{1}{\sqrt{2}}-\frac{\sqrt{5+4\sqrt{2}}}{2}, \nonumber\\
\beta_4&=\frac{1}{\sqrt{2}}+\frac{\sqrt{5+4\sqrt{2}}}{2} 
\end{align}
of $R_2(\beta)=0$. We see, however, that only the roots $\beta_1$ and $\beta_2$ give rise to the common root $(\mu, \nu)$ for Eq. (\ref{eqn:fi}). For both $\beta_1$ and $\beta_2$, the common root $(\mu,\nu)$ is given by (1, -1). As was stated below Eq. (\ref{eqn:resultant}), this is because Eq. (\ref{eqn:resultant}) is a necessary but not a sufficient condition for Eq. (\ref{eqn:fi}) to possess a common root $(\mu,\nu)$.

\subsection{\label{subsec:constraints}Constraints for $\sqrt{\bm{a}^2}$, $\sqrt{\bm{b}^2}$ and $\sqrt{\bm{c}^2}$\protect\\}

In order to obtain $\ba$, $\bb$ and $\bc$, we must keep up the inequalities
\begin{align}
&-ab\leqq \ba\cdot\bb\leqq ab, \nonumber\\
&-bc\leqq \bb\cdot\bc\leqq bc, \label{eqn:abcp}\\
&-ca\leqq \bc\cdot\ba\leqq ca, \nonumber
\end{align}
where $a$, $b$ and $c$ denote $\sqrt{\ba^2}$, $\sqrt{\bb^2}$ and $\sqrt{\bc^2}$, respectively. With the help of the inequality $\wp(\omega)\geqq e_1$, which is valid in the case noted around Eq. (\ref{eqn:wpinequality}), we easily obtain the maxima and the minima of $\ba\cdot\bb$, $\bb\cdot\bc$ and $\bc\cdot\ba$. They depend on the signatures of $(\alpha+\beta+2)z_1$, $\mu z_1$ and $-\nu z_1$. For the above example $\alpha=\beta=\beta_1$, $\mu=1$, $\nu=-1$, we have $(\alpha+\beta+2)z_1>0$, $\mu z_1=-\nu z_1<0$ and obtain the condition
\begin{align} 
\mbox{min}(X,Y,Z)\geqq e_2> e_3\geqq\mbox{max}(U,V,W), 
\label{eqn:MM1}
\end{align}
where $X$, $Y$, $Z$, $U$, $V$ and $W$ are given by 
\begin{align}
X&=\frac{1}{z_1}\biggl(\frac{ab-d_1}{\alpha+\beta+2}-z_2\biggr),\nonumber \\
Y&=-\frac{1}{z_1}(bc+d_2+z_2), \nonumber\\
Z&=-\frac{1}{z_1}(ca+d_3+z_2), \nonumber\\
U&=-\frac{1}{z_1}\biggl(\frac{ab+d_1}{\alpha+\beta+2}+z_2\biggr), \label{eqn:XYZ}\\
V&=\frac{1}{z_1}(bc-d_2-z_2), \nonumber\\
W&=\frac{1}{z_1}(ca-d_3-z_2). \nonumber
\end{align}
Sufficient conditions for (\ref{eqn:MM1}) to hold valid are 
\begin{align}
&f(U),f(V),f(W)< 0, \nonumber\\
&f(\text{min}(X,Y,Z))< 0, \label{eqn:AB}\\
&\text{max}(U,V,W)<u_{-}<\text{min}(X,Y,Z), \nonumber
\end{align}
where $f(u)$ and $u_{-}$ are defined by
\begin{align}
f(u)&=4(u-e_1)(u-e_2)(u-e_3)\equiv 4u^3-g_2 u-g_3
\end{align}
and $u_{-}=-\frac{\sqrt{g_2}}{2\sqrt{3}}$, satisfying $f'(u_{-})=0$. It can be numerically checked that, e.g. $\sqrt{\ba^2}=6$, $\sqrt{\bb^2}=3$, $\sqrt{\bc^2}=1$ satisfy Eq. (\ref{eqn:AB}). Thus we realize that there indeed exist solutions that we intended to obtain.  

As for the case $\alpha=\beta=\beta_2$, $\mu=1$, $\nu=-1$, we have $(\alpha+\beta+2)z_1>0$, $\mu z_1=-\nu z_1>0$. We then find that
\begin{align}
\text{min}(X,V,W)\geqq e_2>e_3\geqq\text{max}(U,Y,Z)
\end{align}
should be satisfied. In this case, we obtain the sufficient conditions
\begin{align}
&f(U),f(Y),f(Z)< 0,\nonumber\\
&f(\text{min}(X,V,W))< 0,\nonumber\\
&\text{max}(U,Y,Z)<u_{-}<\text{min}(X,V,W).
\end{align}
We find that, e.g. $\sqrt{\bm{a}^2}=\sqrt{5/6}$, $\sqrt{\bm{b}^2}=1$ and $\sqrt{\bm{c}^2}=2$ are the case.

\subsection{\label{expression}The expressions for $\bm{a}$, $\bm{b}$, and  $\bm{c}$\protect\\}

We finally discuss how to obtain $\ba$,  $\bb$, $\bc$ from given $\ba^2$, $\bb^2$, $\bc^2$, $\ba\cdot\bb$, $\bb\cdot\bc$ and $\bc\cdot\ba$. Expressing  $\ba\cdot\bb$, $\bb\cdot\bc$ and $\bc\cdot\ba$ as
\begin{align}
\ba\cdot\bb&=ab\hspace{0.5mm}\cos\theta(\omega), \nonumber\\
\bb\cdot\bc&=bc\hspace{0.5mm}\cos\phi(\omega), \label{eqn:bcp}\\
\bc\cdot\ba&=ac\hspace{0.5mm}\cos\psi(\omega), \nonumber
\end{align}
we introduce the unit vectors $\be(\omega)$ and $\bff(\omega)$ as
\begin{align}
&\be(\omega)=\left(\sin\theta(\omega)\cos\alpha_1(\omega),\sin\theta(\omega)\sin\alpha_1(\omega),\cos\theta(\omega)\right), \nonumber\\
&\bff(\omega)=\left(\sin\psi(\omega)\cos\alpha_2(\omega),\sin\psi(\omega)\sin\alpha_2(\omega),\cos\psi(\omega)\right).
\end{align} 
Comparing Eq. (\ref{eqn:bcp}) with Eq. (\ref{eqn:ab}), we can regard $\theta(\omega), \phi(\omega)$ and $\psi(\omega)$ as given by $K(\omega)$. The angles $\alpha_1(\omega)$ and $\alpha_2(\omega)$ can be chosen so as to satisfy
\begin{align}
\text{cos}\phi(\omega)&=\text{cos}\theta(\omega)\text{cos}\psi(\omega)+\text{sin}\theta(\omega)\text{sin}\psi(\omega)\text{cos}\alpha(\omega),\\
\alpha(\omega)&=\alpha_1(\omega)-\alpha_2(\omega).
\end{align}
We also introduce the unit vectors $\bi(\omega, \omega^{\prime})$, $\bj(\omega, \omega^{\prime})$ and $\bk(\omega, \omega^{\prime})$. They can be assumed to satisfy 
\begin{align}
\bi(\omega, \omega^{\prime})\cdot\bj(\omega, \omega^{\prime})&=0,\nonumber\\
\bi(\omega, \omega^{\prime})\cdot\bk(\omega, \omega^{\prime})&=0,\\
\bj(\omega, \omega^{\prime})\cdot\bk(\omega, \omega^{\prime})&=\cos\alpha(\omega). \nonumber
\end{align}
Then we set
\begin{align}
\ba(\omega, \omega^{\prime})&=a\bi(\omega, \omega^{\prime}), \nonumber\\
\bb(\omega, \omega^{\prime})&=b\cos\theta(\omega)\bi(\omega, \omega^{\prime})+b\sin\theta(\omega)\bi(\omega, \omega^{\prime})\times\bj(\omega, \omega^{\prime}), \nonumber\\
\bc(\omega, \omega^{\prime})&=c\cos\psi(\omega)\bi(\omega, \omega^{\prime})+c\sin\psi(\omega)\bi(\omega, \omega^{\prime})\times\bk(\omega, \omega^{\prime}).
\end{align}
It is clear that the above  $\ba(\omega, \omega^{\prime})$, $\bb(\omega, \omega^{\prime})$ and $\bc(\omega, \omega^{\prime})$ satisfy (\ref{eqn:ab}). We find that $\ba\cdot(\bb\times\bc)$ is calculated to be $abc\hspace{0.5mm}\sin\theta\sin\psi\cos\alpha$. The relationship among $\ba$, $\bb$, $\bc$, $\be$, $\bff$, $\bi$, $\bj$ and $\bk$ can be depicted as in Fig.1. We note that we still have freedom in choosing $\bi(\omega, \omega^{\prime})$ and $\bj(\omega, \omega^{\prime})$.

\begin{figure}
\includegraphics[width=6cm]{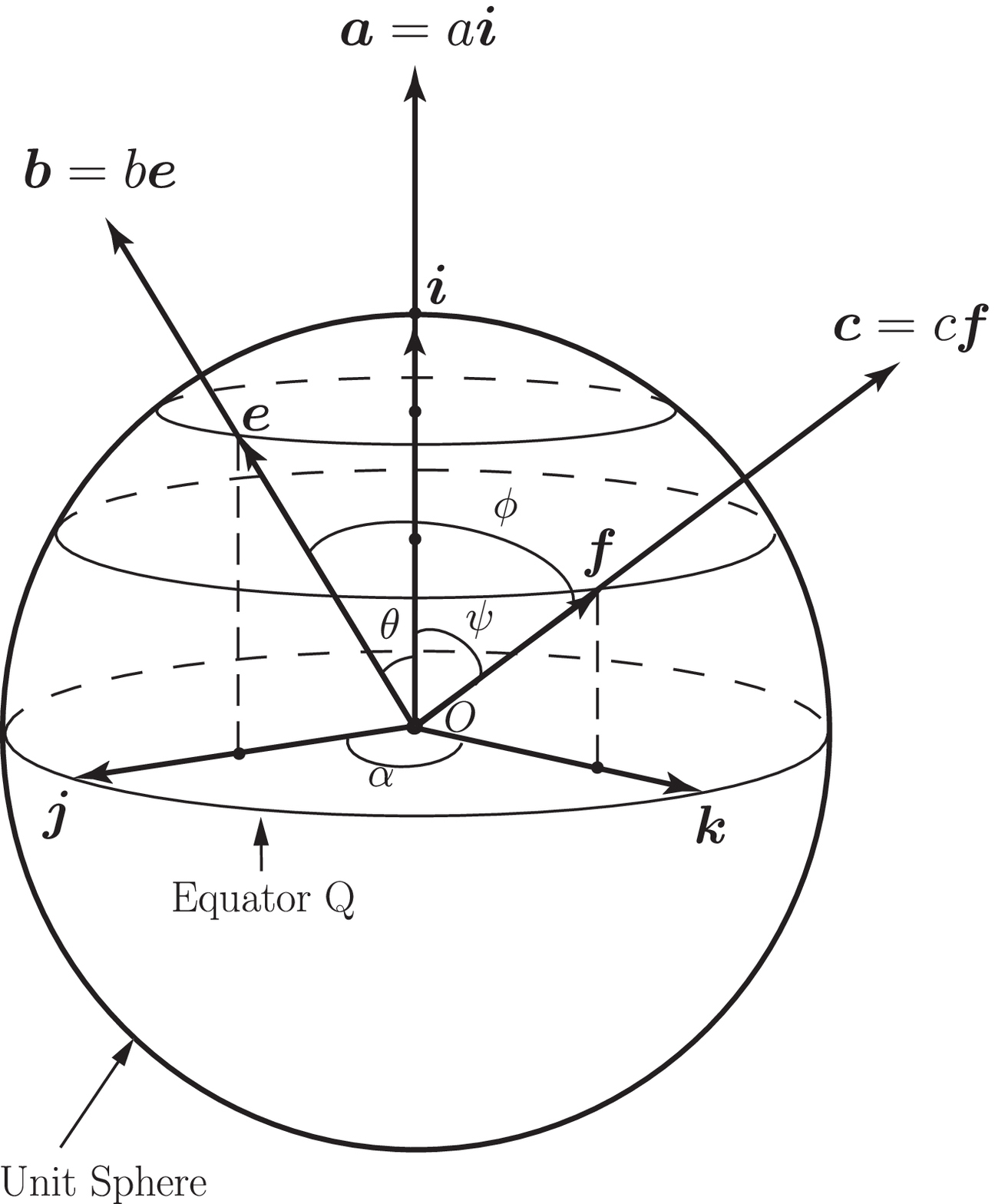} 
\caption{Relationship among $\bm{a}$, $\bm{b}$, $\bm{c}$, $\bm{e}$, $\bm{f}$, $\bm{i}$, $\bm{j}$, $\bm{k}$.}
\end{figure}
\newpage

\section{\label{sec:Summary And Discussion}SUMMARY AND DISCUSSION \protect\\}

We have obtained solutions of the Skyrme model of the form $g(x)=h(k\cdot x, l\cdot x, m\cdot x)$ with the momenta $k, l, m$ satisfying $k^2=l^2=m^2=0$. We found that the variables $\xi, \eta$ and $\zeta$ defined by Eq. (\ref{eqn:xi}) and the fields $\ba, \bb$ and $\bc$ defined by Eq. (\ref{eqn:aia}) are convenient to describe the problem. The existence condition for $g(x)$ and the field equation were expressed in compact forms as seen in Eqs. (\ref{eqn:condition}) and (\ref{eqn:EQMotion1}). To solve these equations, we introduced the {\it Ans\"atze} (\ref{eqn:ansatz1}) and (\ref{eqn:ansatz2}) for the derivatives of $\ba, \bb$ and $\bc$ with respect to $\xi, \eta$ and $\zeta$. The {\it Ans\"atze} contained nine parameters $\alpha, \beta, \gamma, \mu, \nu, \kappa, \lambda, \rho$ and $\sigma$. It was observed that the consistency of the {\it Ans\"atze} yield five constraints (\ref{eqn:constraints5}) for the parameters, leaving $\alpha, \beta, \mu$ and $\nu$ independent. Through the {\it Ans\"atze}, we found that $\ba^2, \bb^2$ and $\bc^2$ are constants, $\ba\cdot\bb, \bb\cdot\bc$ and $\bc\cdot\ba$ are given by the Weierstrass $\wp$ function, and $\ba\cdot(\bb\times\bc)$ is given by the derivative of the $\wp$ function.

With the help of the {\it Ans\"atze}, the field equation was reduced to six algebraic constraints for the parameters $\alpha, \beta, \mu, \nu, d_1, d_2$ and $d_3$, where $d_1, d_2$ and $d_3$ are arbitrary constants contained in $\ba\cdot\bb, \bb\cdot\bc$ and $\bc\cdot\ba$, respectively. Three of the six constraints were used to fix $d_1, d_2$ and $d_3$. The structure of the remaining three constraints for $\alpha, \beta, \mu$ and $\nu$ was discussed. It was confirmed numerically that there indeed exist solutions of the field equation.

From the solutions $\ba, \bb$ and $\bc$, we can construct $A_{\mu}^{\alpha}(x)$ by Eq. (\ref{eqn:Amu}). The matrix field $g(x)$ is obtained as
\begin{align}
g(x)=g(x_0)\overline{P}_{\lambda}\exp\left[i\int^1_0d\lambda A_{\mu}^{\alpha}\left[x(\lambda)\right]\frac{dx^{\mu}(\lambda)}{d\lambda}\tau^{\alpha}\right],
\end{align}
where $x(\lambda)$ parametrizes a path from $x_0=x(0)$ to $x=x(1)$, and $\overline{P}_{\lambda}$ is the anti-$\lambda$ ordering operator. The value of $g(x)$ is independent of the choice of the path because of Eq. (\ref{eqn:Amu}). The above expression for $g(x)$ is rather symbolical because the path-ordered quantity is not easy to evaluate explicitly. Some important physical quantities, however, can be calculated in terms of $A_{\mu}^{\alpha}(x)$. One example is the baryon number current discussed in Sec.I and Sec.III. The isospin current touched upon in Sec.I is also calculated through $A_{\mu}^{\alpha}(x)$. Another physical quantity which can be calculated from $A_{\mu}^{\alpha}(x)$ is the energy-momentum tensor $T_{\mu\nu}(x)$ defined by
\begin{align} T_{\mu\nu}=\frac{\partial{\cal L}_S}{\partial A^{\alpha, \mu}}A^{\alpha}_{\nu}-\eta_{\mu\nu}{\cal L}_S.
\end{align}
It is straightforward to obtain a result such as
\begin{align}
T_{\mu}^{\hspace{2mm}\mu}=-\frac{32c_2^2}{c_4}\left[(\ba\cdot\bb)+(\bb\cdot\bc)+(\bc\cdot\ba)\right].
\end{align}
For the solution of this paper, we see that $T_{\mu}^{\hspace{2mm}\mu}$ is given as a linear function of $\wp(\omega+\omega_3)$. We can also obtain
\begin{align}
k^{\mu}k^{\nu}T_{\mu\nu}=16\kappa_1^2\frac{c_2^3}{c_4^2}\left[(\bb+\bc)^2+2(\bb\times\bc)^2\right]
\end{align}
and find that it contains $[\wp(\omega+\omega_3)]^2$. Up to now, we have been considering lightlike momenta $k, l$ and $m$. We note that $t^{\mu}s^{\nu}T_{\mu\nu}$ should be non-negative for all pairs of future directed timelike vectors $t$ and $l$ \cite{Gibbons}. 

The solutions obtained in this paper are of wave character. They are regarded as the superpositions of three plane waves. We end this paper by asking what the {\it Ansatz} leading to the exact solitonic solutions of the Skyrme model is.


\begin{acknowledgments}
The authors thank Shinji Hamamoto, Takeshi Kurimoto, Hitoshi Yamakoshi, Hiroshi Kakuhata, Kouichi Toda, Masataka Ueno, Makoto Nakamura, and Hideaki Hayakawa for discussions. One of the authors (M.H) is grateful to Noriaki Setoh and Takashi Sugatani for valuable comments. This work was supported in part by a Japanese Grant-in-Aid for Scientific Research from the Ministry of Education, Culture, Sports, Science and Technology (No.13135211).
\end{acknowledgments}

\end{document}